\documentclass[aps,pra,reprint,superscriptaddress,showpacs]{revtex4-2}
\usepackage{graphicx,amsmath,amssymb,ulem}
\usepackage{verbatim}
\usepackage{ragged2e}
\usepackage{color}
\usepackage{xcolor}
\usepackage{braket}
\usepackage{amsmath}
\usepackage{hyperref}
\usepackage{amssymb}
\usepackage[utf8]{inputenc}

\newcommand{\mydoi}[2]{\href{https://doi.org/#1}{#2}}

\begin{document}
\title{Steady states of \ensuremath{\Lambda}-type three-level systems excited by quantum light in lossy cavities}


\author{H.~Rose}
\affiliation{Paderborn University, Department of Physics, Warburger Stra\ss{}e 100, D-33098 Paderborn, Germany\looseness=-1}
\
\author{O.~V.~Tikhonova}
\affiliation{Faculty of Physics, Moscow State University, Leninskie Gory, 1, Moscow, 119991 Russia}
\
\author{T.~Meier}
\affiliation{Paderborn University, Department of Physics, Warburger Stra\ss{}e 100, D-33098 Paderborn, Germany\looseness=-1}
\
\author{P.~R.~Sharapova}
\affiliation{Paderborn University, Department of Physics, Warburger Stra\ss{}e 100, D-33098 Paderborn, Germany\looseness=-1}
\


\begin{abstract}
The interaction between quantum light and matter is being intensively studied for systems that are enclosed in high-$Q$ cavities which strongly enhance the light-matter coupling. However, for many applications, cavities with lower $Q$-factors are preferred due to the increased spectral width of the cavity mode. Here, we investigate the interaction between quantum light and matter represented by a $\Lambda$-type three-level system in lossy cavities, assuming that cavity losses are the dominant loss mechanism.  We demonstrate that cavity losses lead to non-trivial steady states of the electronic occupations that can be controlled by the loss rate and the initial statistics of the quantum fields. The mechanism of formation of such steady states can be understood on the basis of the equations of motion. Analytical expressions for steady states and their numerical simulations are presented and discussed. 
\end{abstract}

\maketitle

\section{INTRODUCTION}

The quality factor ($Q$-factor) of optical cavities is a key property when studying light-matter interaction inside cavity systems, since it provides information about the lifetime of cavity-photons, and consequently about the width of their frequency distribution. While high-$Q$ cavities are of great interest due to their enhancement of light-matter interaction \cite{Nat425.2003,OE22.2014,NatCommun5.2014,PRL95.067401}, low-$Q$ cavities are applied when a broader distribution in frequency space is advantageous, e.g., for coupling to the resonances of inhomogeneously broadened systems \cite{PRL95.2005,PRB92.2015,ComPhys2020}. Furthermore, low-$Q$ cavities are of interest for quantum-information processing \cite{PRA79.2009,PRA82.2010,OE29.2021} and recently attract more attention due to the formation of new quasinormal modes \cite{PRL122.2019}.

$\Lambda$-type three-level systems (3LS) exhibit unique properties that are beneficial in a wide range of quantum applications \cite{qmemnat,qmemreview,qmemhybrid,QRepeater,OStorage,ONN,nature04353}, and show interesting effects in the presence of radiative losses \cite{Rose2021}. Optically-excited semiconductor quantum wells in microcavities have been demonstrated to form a $\Lambda$-system from two exciton-polariton states and the $2p$-exciton state, which can be resonantly driven with short terahertz pulses \cite{PRL108,klettke2013}. However, there are no high-$Q$ cavities that efficiently support terahertz radiation, which is why lossy cavities are a suitable description for such kind of systems \cite{THz}. Another relevant realization of $\Lambda$-systems is based on negatively charged InAs quantum dots placed in a magnetic field \cite{nphys1054}. Such systems usually show relaxation times in the range of nanoseconds, which is much longer than the lifetime of photons in the case of a bad cavity, making cavity losses the dominant loss mechanism in the system.

In this article, we highlight an interesting effect which appears when $\Lambda$-type 3LS are embedded in lossy cavities whose decay rates are comparable to or exceed the light-matter coupling. To study the transition from the lossless case to lossy cavities, also smaller loss rates are considered. 
After explaining the theoretical model in Section~\ref{sec:thmod}, we present the main results and their discussion in Section~\ref{sec:results}, and we end with a conclusion in Section~\ref{sec:conclusion}.


\section{THEORETICAL MODEL}\label{sec:thmod}

\begin{figure}[ht]
	\centering
	\includegraphics[width=0.45\textwidth]{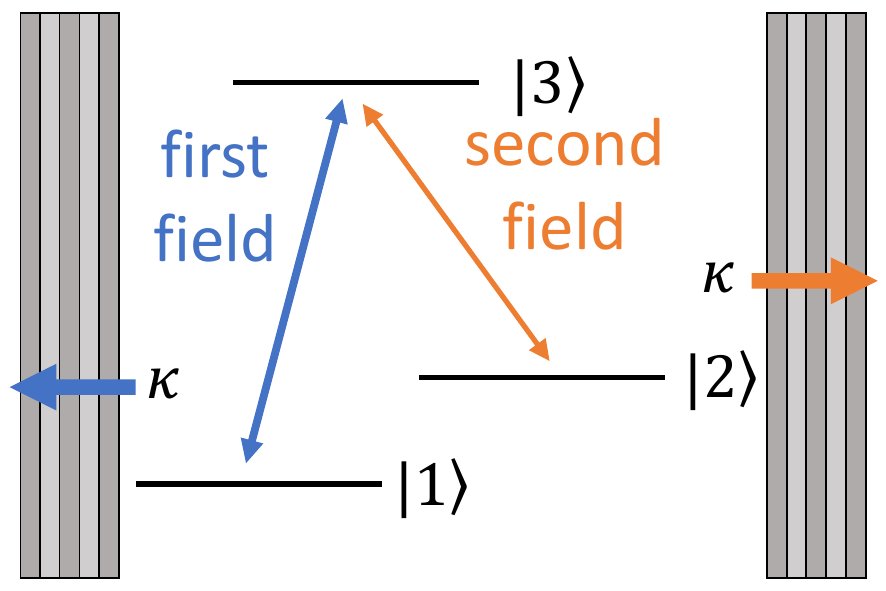}
	\caption{Schematical illustration of a $\Lambda$-type 3LS in a two-mode cavity with losses.}
	\label{fig:3LS}
\end{figure}

A $\Lambda$-type 3LS enclosed in a cavity, schematically shown in Figure~\ref{fig:3LS}, is described with the following Hamiltonian $\hat{H}$, after the rotating-wave approximation was applied:
\begin{subequations}
\begin{align}
\hat{H} &= \hat{H}_0 + \hat{H}_\mathrm{I},\\
\hat{H}_0 &= \sum_{n=1}^3 E_n \hat{\sigma}_{nn} + \sum_{j=1}^2 \hbar \omega_j \left( \hat{a}^\dagger_j \hat{a}_j + \frac{1}{2} \right),\\
\hat{H}_\mathrm{I} &= \sum_{j=1}^2 g_j (\hat{a}^\dagger_j \hat{\sigma}_{j3} + \hat{a}_j \hat{\sigma}_{3j}).
\end{align}
\end{subequations}

Here, the Hamiltonian $\hat{H}$ is separated into $\hat{H}_0$, which describes the free energy of the electronic states and the light fields, and $\hat{H}_{\mathrm{I}}$, describing the interaction between light and matter. $E_n$ denotes the energy of the $n$-th electronic level, $\hat{\sigma}_{ij}=\ket{i}\bra{j}$ is the transition operator for electronic states, $\omega_j$ is the frequency of the $j$-th cavity mode, $\hat{a}^\dagger_j (\hat{a}_j)$ is the bosonic creation (annihilation) operator for a photon of the $j$-th cavity mode, and $g_j$ is the coupling strength between the electronic system and the photons from the $j$-th cavity mode. We will proceed with $g_1 = g_2 = g$, i.e., considering for simplicity the same coupling strength for both transitions.

The dynamics of the system is fully described by its time-dependent density matrix (DM) $\hat{\rho}$ which we consider in the interaction picture. After including a cavity loss rate $\kappa$, which for simplicity is taken to be identical for both modes, by a Lindblad term \cite{Lindblad1976},  the equation of motion reads:
\begin{subequations}
	\begin{align}
	\frac{\partial}{\partial t} \hat{\rho}_\mathrm{I} &= \frac{1}{i\hbar} \left[\hat{H}_\mathrm{I}, \hat{\rho}_\mathrm{I}\right] + \kappa \sum_{j=1}^2 \mathcal{L}(\hat{a}_j),\label{eq:2a}\\
	\mathcal{L}(\hat{X}) &= \hat{X}  \hat{\rho}_\mathrm{I} \hat{X}^\dagger - \frac{1}{2} \left\{ \hat{X}^\dagger \hat{X}  \hat{\rho}_\mathrm{I}\right\},
	\end{align}
\end{subequations}
where the square brackets denote the commutator, and the curly brackets denote the anti-commutator. We furthermore introduce the following rescalings:
\begin{align}
\tilde{\kappa} = \kappa \frac{\hbar}{g},\quad \tilde{t} = t \frac{g}{\hbar},
\end{align}
where $\tilde{\kappa}$ and $\tilde{t}$ are dimensionless quantities that measure the cavity loss rate $\kappa$ in units of $\frac{g}{\hbar}$ and the time $t$ in units of $\frac{\hbar}{g}$. This allows us to transform Eq.~(\ref{eq:2a}) into:
\begin{align}
\frac{\partial}{\partial \tilde{t}} \hat{\rho}_\mathrm{I} &= i \sum_{j=1}^2 \left[ (\hat{a}^\dagger_j \hat{\sigma}_{j3} + \hat{a}_j \hat{\sigma}_{3j}), \hat{\rho}_\mathrm{I}\right] + \tilde{\kappa} \sum_{j=1}^2 \mathcal{L}(\hat{a}_j).
\end{align}
For a more transparent analysis of the DM, we study its elements with the following decomposition
\begin{align}
\hat{\rho}_\mathrm{I} &= \sum_{\substack{n=1 \\ n'=1}}^3 \sum_{\substack{k=0,k'=0 \\ m=0,m'=0}}^{\infty} p_{\substack{n,k,m \\ n',k',m'}} \ket{n,k,m}\bra{n',k',m'}, 
\end{align}
where the basis $\ket{n,k,m}$ is composed of three subsystems, namely, the electronic state $\ket{n}$ and the Fock states $\ket{m}$ and $\ket{k}$ of the first and the second field, respectively. This decomposition allows us to formulate  equations for individual DM elements, where we consider no optical detuning between the cavity photons and the electronic transition:
\begin{widetext}
	\begin{align}
	\begin{split}
	\frac{\partial}{\partial \tilde{t}} p_{\substack{n,k,m \\ n',k',m'}}\left(\tilde{t}\right)=i\bigg(&p_{\substack{n+2,k-1,m \\ n',k',m'}}\left(\tilde{t}\right)\sqrt{k}+p_{\substack{n-2,k+1,m \\ n',k',m'}}\left(\tilde{t}\right)\sqrt{k+1} -p_{\substack{n,k,m \\ n'+2,k'-1,m'}}\left(\tilde{t}\right)\sqrt{k'}-p_{\substack{n,k,m \\ n'-2,k'+1,m'}}\left(\tilde{t}\right)\sqrt{k'+1}\bigg)\\
	+i\bigg(&p_{\substack{n+1,k,m-1 \\ n',k',m'}}\left(\tilde{t}\right)(1-\delta_{n+1,2})\sqrt{m}+p_{\substack{n-1,k,m+1 \\ n',k',m'}}\left(\tilde{t}\right)(1-\delta_{n-1,1})\sqrt{m+1}\\
	&-p_{\substack{n,k,m \\ n'+1,k',m'-1}}\left(\tilde{t}\right)(1-\delta_{n'+1,2})\sqrt{m'}-p_{\substack{n,k,m \\ n'-1,k',m'+1}}\left(\tilde{t}\right)(1-\delta_{n'-1,1})\sqrt{m'+1}\bigg)\\
	+\tilde{\kappa}\bigg[&p_{\substack{n,k+1,m \\ n',k'+1,m'}}\left(\tilde{t}\right)\sqrt{k+1}\sqrt{k'+1}+p_{\substack{n,k,m+1 \\ n',k',m'+1}}\left(\tilde{t}\right)\sqrt{m+1}\sqrt{m'+1}-\frac{1}{2}p_{\substack{n,k,m \\ n',k',m'}}\left(\tilde{t}\right)\big(k+k'+m+m'\big)\bigg].
	\label{eq:Motion}
	\end{split}
	\end{align}
\end{widetext}
Here, the Kronecker delta takes into account the dipole-forbidden transition between the electronic levels $\ket{1}$ and $\ket{2}$. The population dynamics of the state $\ket{n}$ can be calculated as follows:
\begin{align}
O_n = \sum_{k=0}^\infty \sum_{m=0}^\infty p_{\substack{n,k,m \\ n,k,m}}.
\end{align}

At the initial moment of time, we consider our system in the electronic ground state, so that the energetically lowest level $\ket{1}$ is occupied, whereas the remaining levels are empty. Therefore, only DM elements with $n=n'=1$ are non-zero initially. As the initial states of light, we consider arbitrary quantum light modes $\ket{\Psi_1}$ and $\ket{\Psi_2}$ for the first and the second field, respectively, which can be expanded into Fock states as follows:
\begin{subequations}
	\begin{align}
	\ket{\Psi_1} &= \sum_{k=0}^\infty c_k \ket{k},\\
	\ket{\Psi_2} &= \sum_{m=0}^\infty c'_m \ket{m}.
	\end{align}
\end{subequations}
Thus, the initial non-vanishing elements of the DM are given by
\begin{align}
p_{\substack{1,k,m \\ 1,k',m'}}(\tilde{t} = 0) = c_k c'_m c^*_k c'^*_m.
\end{align}
The quantum light that is considered in the scope of this work includes coherent states $\ket{\alpha}$, squeezed vacuum states $\ket{\xi}$, and Fock states $\ket{n}$. For these states, the photon statistics and the mean photon numbers $\braket{\hat{n}}=\braket{\hat{a}^\dagger \hat{a}}$ are given by
\begin{subequations}
	\begin{align}
	&\ket{\alpha} = \sum_{k=0}^\infty e^{\frac{-|\alpha|^2}{2}} \frac{\alpha^k}{\sqrt{k!}} \ket{k},\\
	&\ket{\xi} = \sum_{k=0}^\infty  (-1)^k \frac{\sqrt{(2m)!}}{2^m m!} \frac{(e^{i\vartheta} \tanh(r))^k}{\sqrt{\cosh(r)}} \ket{2k},\\
	&\ket{n} = \sum_{k=0}^\infty \delta_{n,k} \ket{k},\\
	&\bra{\alpha} \hat{n} \ket{\alpha} = |\alpha|^2,\\
	&\bra{\xi} \hat{n} \ket{\xi} = \sinh^2(r),\\
	&\bra{n} \hat{n} \ket{n} = n,
	\end{align}
\end{subequations}
where the parameter of squeezing was decomposed with $\xi = r e^{i\vartheta}$.

\section{RESULTS AND DISCUSSION}\label{sec:results}

The results presented in this article are divided into four subsections: in Section~\ref{sec:a}, we demonstrate that non-trivial steady states can be created in lossy cavities. This is followed by a general analysis of the system under consideration in Section~\ref{sec:b}, where the origin of steady states is explained from the relationship between DM elements. Subsequently, analytical solutions that allow a deeper understanding of this phenomenon are presented in Section~\ref{sec:ana}. Finally, in Section~\ref{sec:d}, this effect is demonstrated for different photon statistics and different regimes of cavity losses. All numerical solutions shown below were obtained by integrating the equations of motion using the fourth-order Runge-Kutta method.

\subsection{Population dynamics with cavity losses}\label{sec:a}

\begin{figure}[ht]
	\centering
	\includegraphics[width=0.5\textwidth]{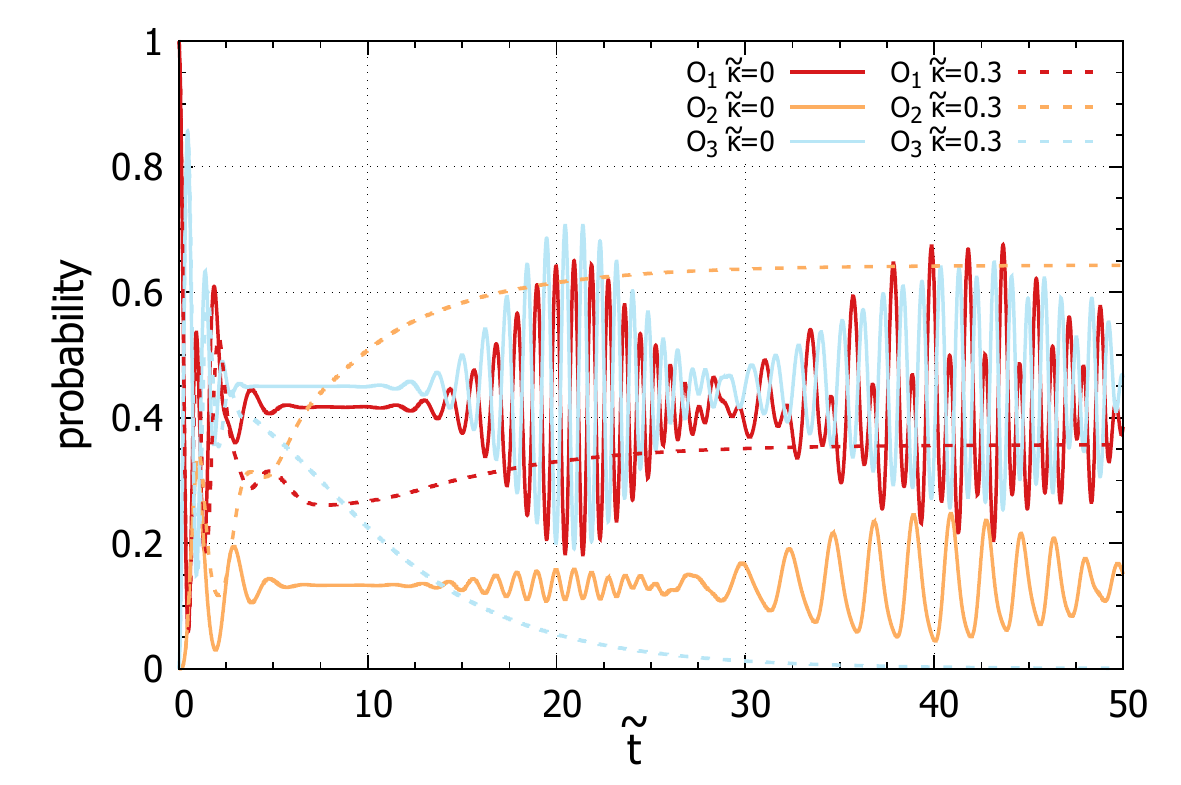}
	\caption{Population dynamics with ($\tilde{\kappa} = 0.3$) and without ($\tilde{\kappa}=0$) cavity losses for a coherent state with a mean photon number of $10$ as first field and the vacuum state as the second field.}
	\label{fig:demo}
\end{figure}

We start our investigation with a demonstration of the population dynamics of the electronic states. This dynamics is shown in Figure~\ref{fig:demo}, where the first field is a coherent state with a mean photon number of $10$ and the second field is a vacuum state. Each dynamics is shown for $\tilde{\kappa}=0$ and $\tilde{\kappa}=0.3$, respectively, allowing a direct comparison. In the lossless case, we find the well-known collapse and revival behavior caused by the superposition of oscillations belonging to different Fock states contained in the coherent state. The population of the second level $O_2$ is rather small, since the second field is a vacuum state, and therefore, the only possibility to transfer electrons to the second level is due to the absorption of a photon from the first field, which promotes an electron to the third level, and a subsequent relaxation of the electron into the second level under the emission of a photon.

This behavior changes in the case of finite cavity losses, when the highest population at large times is found in the second state, while the third level is empty. At first glance, such a result seems unexpected, however, described behavior can be understood by carefully studying the system and the relationship between the DM elements, which is done in the next subsection.

\subsection{General analysis of the system}\label{sec:b}

To get an understanding of the previous demonstration,  we divide the DM elements of Eq.~(\ref{eq:Motion})  into interacting elements (IE) and non-interacting elements (NIE). The element is called NIE when its time derivative is zero in the lossless case, i.e. when $\partial_{\tilde{t}} p_{\substack{n,k,m \\ n',k',m'}} = 0$ for $\tilde{\kappa} = 0$, while all other elements are considered as IE. The NIE that can contribute to the population of levels have either the form $p_{\substack{1,0,m \\ 1,0,m}}$ or $p_{\substack{2,k,0 \\ 2,k,0}}$. In presence of cavity losses $\kappa$, the equations of motion for these NIE are given by:
\begin{align}
\partial_{\tilde{t}} p_{\substack{1,0,m \\ 1,0,m}} = \tilde{\kappa} \Big[&  p_{\substack{1,1,m \\ 1,1,m}} +  p_{\substack{1,0,m+1 \\ 1,0,m+1}} (m+1) - p_{\substack{1,0,m \\ 1,0,m}} m\Big],\\
\partial_{\tilde{t}} p_{\substack{2,k,0 \\ 2,k,0}} = \tilde{\kappa} \Big[&  p_{\substack{2,k,1 \\ 2,k,1}} +  p_{\substack{2,k+1,0 \\ 2,k+1,0}} (k+1) - p_{\substack{2,k,0 \\ 2,k,0}} k \Big].
\end{align}
One can observe that a NIE has always two source terms, an IE and a NIE, both from higher indices. Furthermore, a loss term is proportional to the Fock number of the considered element. Due to cavity losses, an IE will contribute to a NIE, while the NIE themself contribute to NIE of a fewer  Fock state number. 
The elements without further decay can be treated as the steady state population:
\begin{align}
O_{1,\mathrm{st}} = p_{\substack{1,0,0 \\ 1,0,0}}(\tilde{t}\rightarrow\infty),\\
O_{2,\mathrm{st}} = p_{\substack{2,0,0 \\ 2,0,0}}(\tilde{t}\rightarrow\infty).
\end{align}

\begin{figure}[ht]
	\centering
	\includegraphics[width=0.4\textwidth]{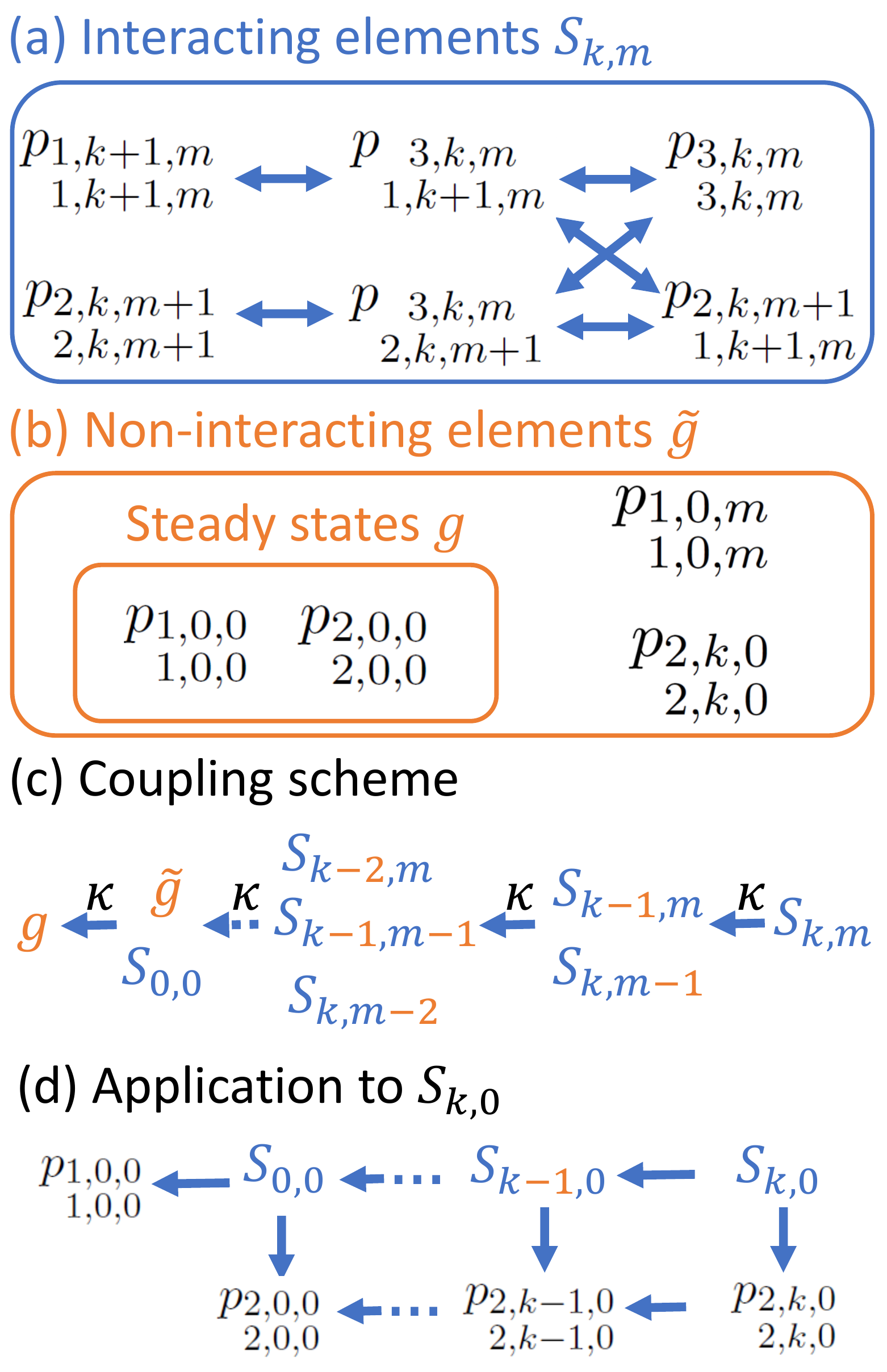}
	\caption{A diagram that illustrates (a) IEs $S_{k,m}$ and their coupling to each other, (b) the NIEs $\tilde{g}$ and the steady states $g$, which is a subset of $\tilde{g}$, (c) the coupling scheme between the DM elements, eventually leading to the formation of the steady state, and (d) the application of this coupling scheme to $S_{k,0}$. $S_{k,0}$ refers to a state in which the first field initially has $k+1$ photons, while the second state is the vacuum state.}
	\label{fig:diagram}
\end{figure}

The steady state population includes the NIE with the lowest field indices for the considered level. It means that all NIE with higher field indices will contribute to this steady state as well. This yields a qualitative understanding of the behavior found in Fig.~\ref{fig:demo}: When the second field is a vacuum state, the only NIE contributing to the population of the first electronic level is $p_{\substack{1,0,0 \\ 1,0,0}}$, since the Fock state index $m$ for the second field must be zero. In contrary, since the first field was chosen as a coherent state with a broad photon number distribution, the NIE $p_{\substack{2,k,0 \\ 2,k,0}}$ is finite as long as there is a finite probability of measuring $k+1$ photons in the first field. Thus, we find more contributions to the steady state population of the second level as compared to the first level. A contribution of different elements to the steady states can be understood from Figure~\ref{fig:diagram}, which presents a diagram illustrating the classification of IEs and NIEs (Fig.~\ref{fig:diagram}(a) and Fig.~\ref{fig:diagram}(b), respectively), as well as the general coupling scheme between these elements in Fig.~\ref{fig:diagram}(c). An application of the general scheme to the example of the vacuum second field is demonstrated in Fig.~\ref{fig:diagram}(d).


Physically, the process can be understood as follows. Due to the excitation of an electron from the first to the third level with a photon of the first field, the electron can then decay to the second level, while emitting a photon belonging to the second field. However, the cavity losses will now destroy this photon, so that the photon cannot excite the electron back to the third level. The probability of such an event is expressed by NIEs. The timescale of a photon destruction  is determined by the value of $\kappa$, namely, depending on $ \kappa $, various steady state populations are formed, which will be discussed later in Sec.~\ref{sec:d}. Note that more precisely the loss rate that acts on the second field is of importance for the formation of the steady state population of the second level, however, we neglect this additional degree of freedom for simplicity.

\subsection{Analytical approach}\label{sec:ana}

In this subsection, we apply the explained above scheme to explicit examples, assuming a single photon or two-photon Fock state as the first field, while the second field remains in the vacuum state. It should be noted that none of these systems can be solved  exactly analytically, since the amount of coupled differential equations is at least six, which in general does not allow for an analytical diagonalization. Therefore, we obtain our analytical solutions by fitting the numerical simulation, while assuring that $\tilde{\kappa}=0$ leads to the exact analytical solution, which was shown in \cite{popolitova2019}. This is realized by solving the problem numerically for different $\tilde{\kappa}$ and determining the analytical functions and coefficients that fit the numerical solution, so that the dependence on $\tilde{\kappa}$ can be identified.

The accuracy of the obtained analytical expressions depends on the considered quantity and the value of $\tilde{\kappa}$, where we find the general trend that the analytical solutions are more accurate for small $\tilde{\kappa}$. Note that $\tilde{\kappa}=0$ does not lead to steady states due to the absence of the loss mechanism, and is therefore not shown in the subsequent figures.

\subsubsection{Single-photon Fock state}

A single photon as the first field leads to a symmetrical behavior for steady states:  After the excitation of the $1-3$ transition, the first field is in a vacuum state, while the absorbed photon may contribute to the second field. Thus, during the dynamics of the system, the first and the second fields are interchanging between a single photon state and a vacuum state. An approximated result for the steady state populations yields:

\begin{align}
O_{1,\mathrm{st}} = \frac{1}{2} + \bigg(\frac{2\tilde{\kappa}^2}{16+4.5\tilde{\kappa}^2}-\frac{0.5\tilde{\kappa}^2}{64+2\tilde{\kappa}^2}\bigg),\label{eq:stsp1}\\
O_{2,\mathrm{st}} = \frac{1}{2} - \bigg(\frac{2\tilde{\kappa}^2}{16+4.5\tilde{\kappa}^2}-\frac{0.5\tilde{\kappa}^2}{64+2\tilde{\kappa}^2}\bigg).\label{eq:stsp2}
\end{align}

\begin{figure}[ht]
	\centering
	\includegraphics[width=0.5\textwidth]{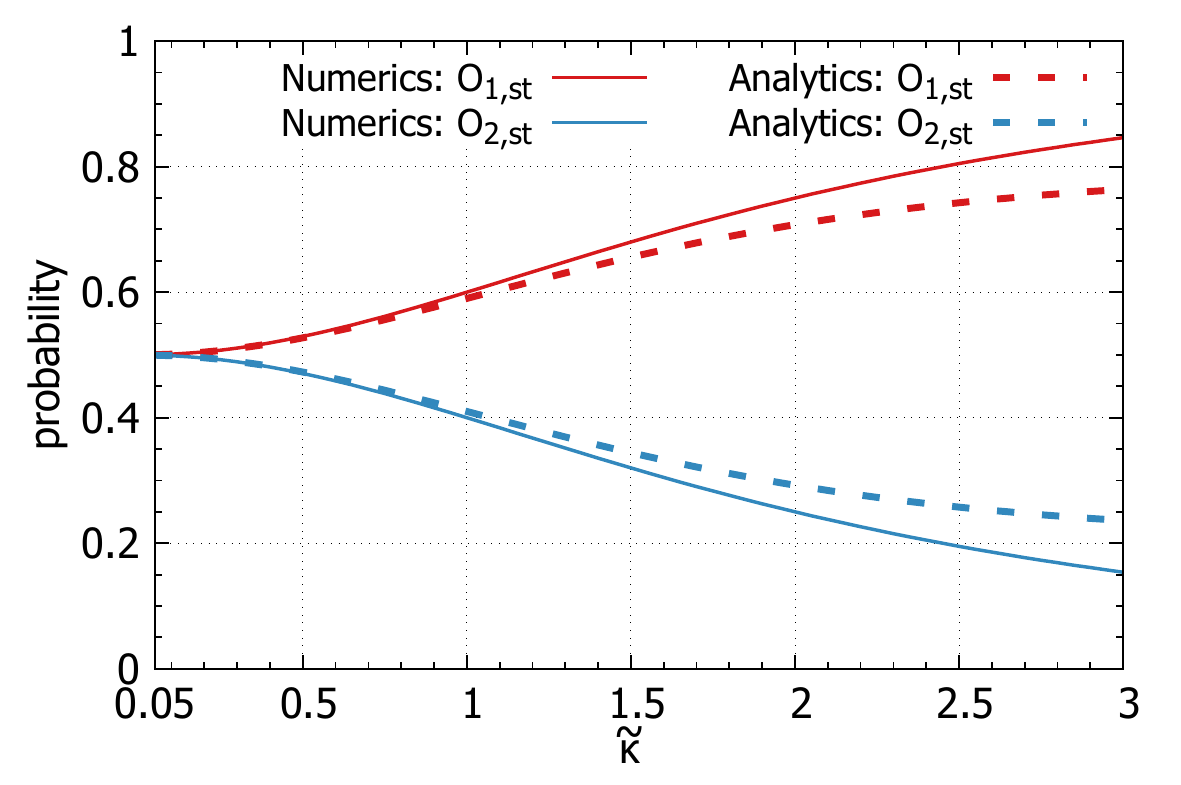}
	\caption{Numerical simulation of $O_{1,\mathrm{st}}$ and $O_{2,\mathrm{st}}$ depending  on $\tilde{\kappa}$ for the case of a single photon state as the first field and a vacuum state as the second field, together with the analytical solutions from Eqs.~(\ref{eq:stsp1}) and (\ref{eq:stsp2}).}
	\label{fig:single}
\end{figure}

Due to the symmetrical behavior, both steady state populations are close to $\frac{1}{2}$ for small $\tilde{\kappa}$, while $O_{1,\mathrm{st}}$ increases with growing $\tilde{\kappa}$, since the photon is initially in the first field, and, therefore, can directly be coupled to the steady state $O_{1,\mathrm{st}}$. We note that there is no NIE of higher indices, since only a single photon is considered. Figure~\ref{fig:single} shows a comparison between the approximate analytical result and the corresponding numerical simulation. One can see that the general trend is reproduced, however, the magnitude of analytical steady states differs from the numerical solution for large $\tilde{\kappa}$.

\subsubsection{Two-photon Fock state}

In contrast to a single photon, when considering a two-photon Fock state as the first field, the NIEs for several field indices are relevant. In this case, the steady state populations are calculated as follows:
\begin{align}
O_{1,\mathrm{st}} &=  p_{\substack{1,0,0 \\ 1,0,0}}(\tilde{t}\rightarrow\infty) = \tilde{\kappa} \int_0^\infty p_{\substack{1,1,0 \\ 1,1,0}}(\tilde{t}')\mathrm{d}\tilde{t}',\\
O_{2,\mathrm{st}} &=  p_{\substack{2,0,0 \\ 2,0,0}}(\tilde{t}\rightarrow\infty) = \tilde{\kappa} \int_0^\infty \Big[p_{\substack{2,0,1 \\ 2,0,1}}(\tilde{t}')+p_{\substack{2,1,0 \\ 2,1,0}}(\tilde{t}')\Big]\mathrm{d}\tilde{t}'.
\end{align}

While $O_{1,\mathrm{st}}$ has an IE as a source term, $O_{2,\mathrm{st}}$ has two source terms originating from an IE and a NIE, we denote their contributions as follows:
\begin{align}
p_N = \tilde{\kappa} \int_0^\infty p_{\substack{2,1,0 \\ 2,1,0}}(\tilde{t})\mathrm{d}\tilde{t},\\
p_I = \tilde{\kappa} \int_0^\infty p_{\substack{2,0,1 \\ 2,0,1}}(\tilde{t})\mathrm{d}\tilde{t}.
\end{align}

\begin{figure}[ht]
	\centering
	\includegraphics[width=0.5\textwidth]{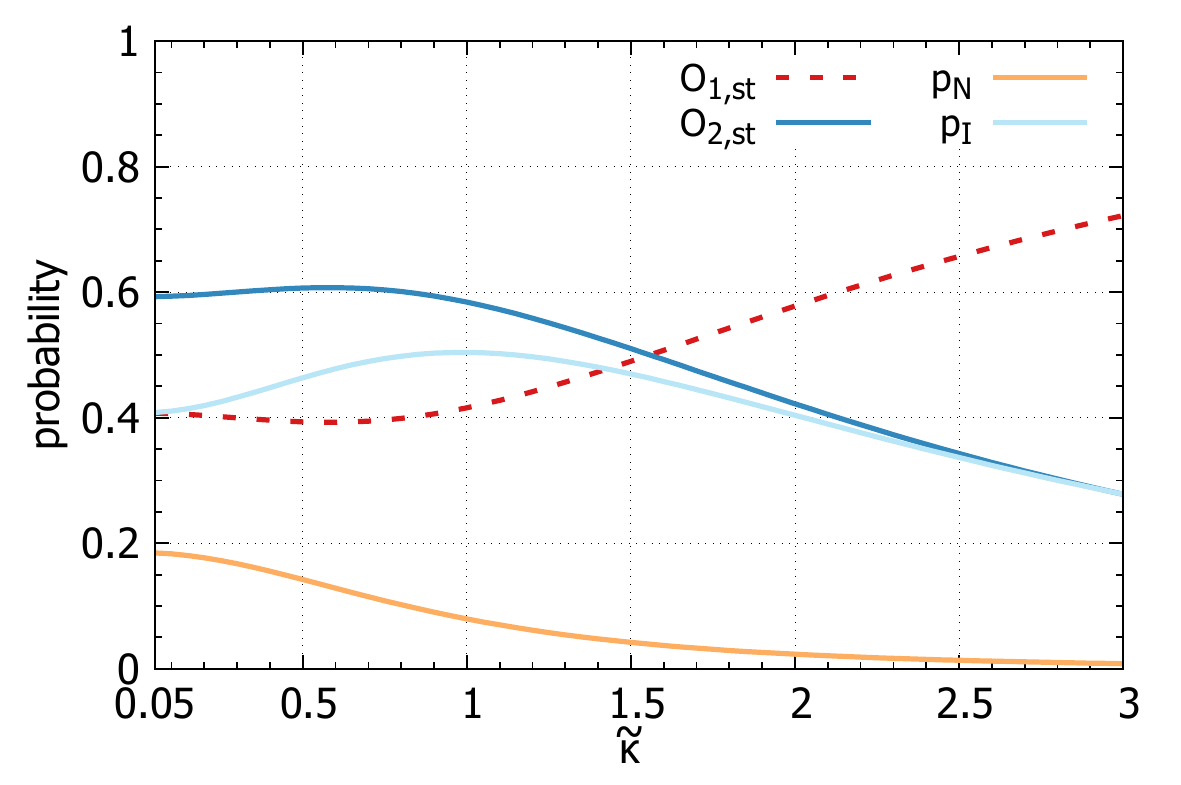}
	\caption{Numerical simulations of $O_{1,\mathrm{st}}$, $O_{2,\mathrm{st}}$, $p_N$, and $p_I$ are shown in dependence on $\tilde{\kappa}$.}
	\label{fig:contr}
\end{figure}

Figure~\ref{fig:contr} shows a numerical simulation of both, the steady states $O_{1,\mathrm{st}}$ and $O_{2,\mathrm{st}}$ as well as the contributions of $p_N$ and $p_I$ in dependence on $\tilde{\kappa}$. One can observe that $p_I$ and $O_{1,\mathrm{st}}$ are nearly at the same value for $\tilde{\kappa}\rightarrow 0$. In contrast to the previous example, however, $O_{2,\mathrm{st}}$ is found to be larger than  $O_{1,\mathrm{st}}$ for small $\tilde{\kappa}$, due to the contribution from $p_N$. For larger values of $\tilde{\kappa}$, the population $O_{1,\mathrm{st}}$ increases, since the cavity losses destroy the photons of the first field faster than the system can perform a cycle of the Rabi oscillation. In the limit for $\tilde{\kappa}\rightarrow\infty$ one would expect the system to remain in its initial state, since the cavity photons are destroyed before the interaction starts.

We proceed the analysis by obtaining an approximate solution for $p_{\substack{2,1,1 \\ 2,1,1}}$, which reads:
\begin{align}
&p_{\substack{2,1,1 \\ 2,1,1}} = \frac{2}{9}\Bigg[\cos\Bigg(\sqrt{3}\tilde{t} \Bigg)e^{-\frac{3}{4}\tilde{\kappa} \tilde{t}}-e^{-\tilde{\kappa} \tilde{t}}\Bigg]^2\notag\\
&+\frac{\tilde{\kappa}}{9\sqrt{3}} \Bigg[ \sin\Bigg(\sqrt{3}\tilde{t}\Bigg)e^{-\frac{7}{4}\tilde{\kappa} \tilde{t}}-\frac{1}{2}\sin\Bigg(2\sqrt{3}\tilde{t}\Bigg)e^{-\frac{3}{2}\tilde{\kappa} \tilde{t}} \Bigg].
\end{align}
This solution is found to be accurate for small $\tilde{\kappa}$, reproduces the exact solution without losses for $\tilde{\kappa}=0$, and is directly connected to $p_N$:
\begin{align}
p_N &= \tilde{\kappa}^2 \int_0^\infty \mathrm{d}\tilde{t} e^{-\tilde{\kappa} \tilde{t}} \int_0^{\tilde{t}} \mathrm{d}\tilde{t}' e^{\tilde{\kappa} \tilde{t}'} p_{\substack{2,1,1 \\ 2,1,1}}(\tilde{t}')\\
&=\frac{\left(960 +32 \tilde{\kappa} ^2-7.75 \tilde{\kappa} ^4\right)}{27
	\left(1.5 \tilde{\kappa} ^2+8\right) \left(24.5 \tilde{\kappa} ^2+24\right)},\quad\mathrm{for}\quad\tilde{\kappa}>0.\label{eq:pN}
\end{align}

\begin{figure}[ht]
	\centering
	\includegraphics[width=0.4\textwidth]{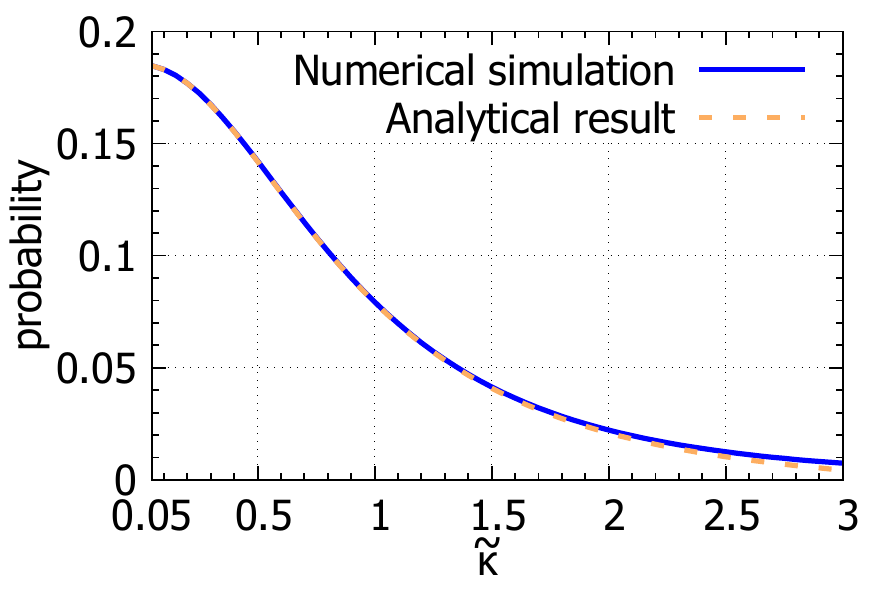}
	\caption{Numerical simulation of $p_N$ depending on $\tilde{\kappa}$ together with the analytical solution from Eq.~(\ref{eq:pN}).}
	\label{fig:pN}
\end{figure}

Figure~\ref{fig:pN} shows the dependence of $p_N$  on $\tilde{\kappa}$ according to Eq.~(\ref{eq:pN}) together with a numerical simulation of $p_N$. It can be seen that both curves overlap, but slowly start to diverge from $\tilde{\kappa} = 1.5$, which verifies the suitability of the presented analytical result. In contrast to $p_N$, it is much more involved to obtain $p_I$, since it results from a larger amount of coupled differential equations, for this reason, we do not show an analytical solution for it.

\subsection{Different initial states of light and loss regimes}\label{sec:d}

In this subsection, numerically simulating  Eq.~(\ref{eq:Motion}), we investigate the steady state population for different initial states of light and for different regimes of $\kappa$: small $\kappa$ ($\tilde{\kappa}\ll 1$), intermediate $\kappa$ ($\tilde{\kappa} \approx 1$), and high $\kappa$ ($\tilde{\kappa}\gg 1$).

\begin{figure}[ht]
	\centering
	\includegraphics[width=0.5\textwidth]{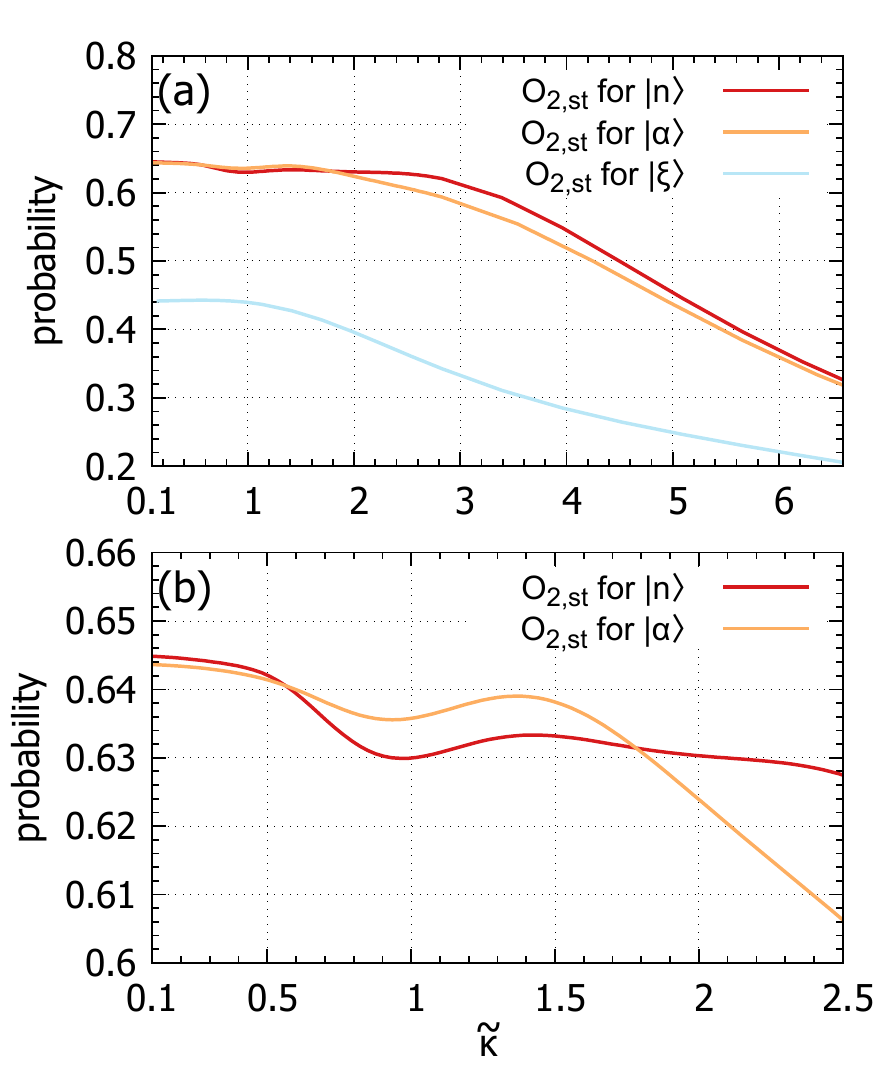}
	\caption{Numerical simulations of $O_{2,\mathrm{st}}$  depending on $\tilde{\kappa}$  for a Fock state $\ket{n}$, a coherent state $\ket{\alpha}$, and a squeezed state $\ket{\xi}$ with a mean photon number of $10$ in all cases. (b) is the zoom of (a) which resolves the curves in the intermediate regime. A dependence of $O_{1,\mathrm{st}}$ on $\tilde{\kappa}$ can be obtained from $O_{1,\mathrm{st}} = 1 -O_{2,\mathrm{st}}$.}
	\label{fig:all10}
\end{figure}

As for the initial states of light, we choose a coherent state $\ket{\alpha}$, a squeezed vacuum state $\ket{\xi}$ and a Fock state $\ket{n}$, all having a mean photon number of $\braket{\hat{n}}=10$, as the first field, while the second field is a vacuum state. Figure~\ref{fig:all10}(a) shows $O_{2,\mathrm{st}}$ for the respective photon statistics in dependence on $\tilde{\kappa}$. One can observe that the squeezed vacuum results in $O_{2,\mathrm{st}}<0.5$ for all $\tilde{\kappa}$. This is caused by a high vacuum component of the squeezed vacuum photon statistics. Such a component does not  initiate an  electron transition, thereby increasing the contribution of the NIE $p_{\substack{1,0,0 \\ 1,0,0}}$. In contrast, this is not the case for the coherent and the Fock states, since these states have a small or vanishing vacuum component. For these states, the NIE contribution to $O_1$ is much less (for small $\kappa$) compared to the squeezed vacuum state, therefore, a higher value of $O_{2,\mathrm{st}}$ is found. Thus, in the range of small $\kappa$, the behavior of steady states is mainly determined by the photon statistics of the respective initial field.

At  high $\kappa$, the steady state $O_{2,\mathrm{st}}$ strives towards zero, since  the first field is being destroyed faster than the occupation can be transferred between electronic levels. This was also found in the case of a single photon and a two-photon Fock state in Sec.~\ref{sec:ana} and remains for different photon statistics. Moreover, one can observe that the steady state population $O_{2,\mathrm{st}}$ decreases faster for a coherent state compared to the Fock state, which is due to a finite probability of measuring fewer-order Fock states in a coherent state, eventually leading to the NIE corresponding to the first level being populated faster. Therefore, even for high $\kappa$ the photon statistics is important, whereas the overall behavior is that the initial electronic state of the system only slightly changes since the light field is destroyed shortly after initialization.

In contrast to small and high $\kappa$, for intermediate $\kappa$ there is no clear trend and even inflection points are present in the steady state population, as can be seen in Fig.~\ref{fig:all10}(b), which is a zoom of Fig.~\ref{fig:all10}(a). To understand the obtained values in this regime, it is advantageous to consider the time-evolution of $O_2(t)$, which consists of Rabi oscillations. A loss rate $\kappa$ of the same order as the light-matter coupling $g$ leads to a damping after one or a few cycles of the Rabi oscillations. 
In this case, the current phase of the Rabi oscillations while the damping occurs is important. E.g. a damping that occurs during a maximum of $O_2$ is more favorable for the higher $O_{2,\mathrm{st}}$ values, and vice versa for a minimum, which leads to the inflection points in the steady state populations. 





\section{CONCLUSION}\label{sec:conclusion}

We demonstrate that the excitation of  a $\Lambda$-type 3LS by quantum light in lossy cavities may lead to non-trivial steady states. These steady states depend on the cavity loss rate and the initial photon statistics of qunatum fields, while their formation can be understood from the coupling scheme of the density matrix elements. To provide a deeper insight into the appearance of the steady states, we present analytical results for the case of a single-photon and a two-photon Fock states. We show the steady state populations for various quantum fields, including Fock, coherent, and squeezed vacuum states of light, and demonstrate a possibility to control and manage the ratio between populations of the electronic levels $\ket{1}$ and $\ket{2}$ due to cavity losses. 


The presented results contribute to a better understanding of highly-relevant cavity systems with low $Q$-factors, which are widely applied in experiments, and improve the insight into the relationship between different density matrix elements in the interaction process.


\section{ACKNOWLEDGMENTS}

The joint grant by the Deutsche Forschungsgemeinschaft (DFG) and the Russian Science Foundation (RSF) (projects SH~1228/2-1, ME~1916/7-1, No.~19-42-04105) is gratefully acknowledged.
We thank the PC$^2$ (Paderborn Center for Parallel Computing) for a computing time grant.

\end{document}